\documentclass[onecolumn,showpacs,preprintnumbers,amsmath,amssymb]{revtex4}
\usepackage{graphicx}
\usepackage{fancyhdr}
\usepackage{dcolumn}
\usepackage{amssymb}
\usepackage{amsmath}
\usepackage[normalem]{ulem}
\usepackage{bm}
\usepackage{textcomp}
  \usepackage{natbib}
  \usepackage[a4paper,left=1.55cm,top=2cm,right=1.55cm]{geometry}

\begin{document}
\preprint{ }
\title{Ground-state properties of light kaonic nuclei signaling symmetry energy at high densities}
\author{Rong-Yao Yang, Si-Na Wei, Wei-Zhou Jiang}
\affiliation{School of Physics, Southeast University, Nanjing  211189, China}

{\begin{abstract}
A sensitive correlation between the ground-state properties of light kaonic nuclei and the symmetry energy at high densities is constructed under the framework of relativistic mean-field theory. Taking oxygen isotopes as an example, we see that a high-density core is produced in kaonic oxygen nuclei, due to the strongly attractive antikaon-nucleon interaction. It is found that the $1S_{1/2}$ state energy in the high-density core of kaonic nuclei can directly probe the variation of the symmetry energy at supranormal nuclear density, and  a sensitive correlation between  the neutron skin thickness and the symmetry energy at supranormal density is established directly. Meanwhile, the sensitivity of the neutron skin thickness to the low-density slope of the symmetry energy is greatly increased in the corresponding kaonic nuclei. These sensitive relationships are established upon the fact that the isovector potential in the central region of kaonic nuclei becomes very sensitive to the variation of the symmetry energy. These findings might provide another perspective to constrain high-density symmetry energy, and await experimental verification in the future.
\end{abstract}}

\maketitle

\section{Introduction}
The determination of symmetry energy is of crucial importance in contemporary nuclear physics, since it can provide vital information on understanding the basic aspects of the strong interaction, such as three-body force~\cite{Xu10a,Tsang12,Steiner12}, tensor force~\cite{Wang17} and short range correlations~\cite{Li15,Hen15,Cai16}. Up to now, the density dependence of  symmetry energy is still not well understood, even though lots of effort has been made to extract the symmetry energy from theory, astrophysical observations and terrestrial experiments~\cite{Chen05,Ozel06,Li08,Tsang09,Carbone10,Xu10b,Steiner10,Zhang10,Tsang12,Steiner12,Zhang13,Li13,Ma13,Guo17,Watts16,Lattimer16}. Though neutron stars (NSs) are natural laboratories for testing the properties of asymmetric nuclear matter, the symmetry energy constrained by NS observation has not reached a satisfactory precision, due to limited data and indistinct observables, e.g., the radii of NSs~\cite{Watts16,Lattimer16}. Meanwhile, constraints on the symmetry energy from heavy-ion collisions suffer from low detection efficiency, large systematic error and transport model dependence~\cite{Li08,Xiao09,Feng10,Ru11}. The properties of finite nuclei are of great use in constraining the symmetry energy near or beneath the normal
density~\cite{Tsang09,Xu10b,Carbone10,Tsang12,Wang10,Liu10,Jiang10,Wang16} owing to the high precision of the structural data of finite nuclear systems. For instance, the symmetry energy and its density slope at normal nuclear density can be determined by the nucleon global optical potentials, which can be extracted from nucleon-nucleus scatterings and the single-particle energy levels of bound states~\cite{Xu10b}. Additionally, a connection between the nuclear symmetry energy at subnormal density and the symmetry energy coefficients of finite nuclei can be constructed from the measured mass of finite nuclei~\cite{Wang10,Liu10,Zhang13,Wang15,Wang16} and the liquid-gas phase transition~\cite{Xu07,Zgh13}. In particular, the neutron skin thickness of neutron-rich nuclei has captured extensive attention, considering that it can serve as a bridge linking the symmetry energy and the structure of NSs~\cite{Horowitz01,Jiang05,Chen10,Zhang13,Zhang15,Jiang15}. After great efforts have been made to measure the root-mean-square (rms) radius of neutron density distribution in finite nuclei~\cite{Klimkiewicz07,Maza11,Abrahamyan12,Horowitz14,Tarbert14}, the symmetry energy near normal nuclear density has been constrained increasingly well. Despite all this progress, inferring the high density properties of symmetry energy from finite nuclei is difficult since the density of a nucleus saturates at normal nuclear density. If a high-density core in a finite nucleus could be created, it would be very interesting to investigate the relationship between the neutron skin thickness of such nuclei and the symmetry energy.

We should look for a unique way to achieve a higher nuclear density in a finite nuclear system, which cannot be obtained just by adding more nucleons. A possible option is to incorporate the strangeness degree of freedom, forming exotic nuclei, such as $\Lambda$ hyperons~\cite{93JS,94JS,06WZJ} and kaon mesons~\cite{99TK,02YA,02TY}. Noticeably, the core density of kaonic nuclei could possibly be as high as two times the normal nuclear density due to the strongly attractive antikaon-nucleon interaction~\cite{02TY}. In fact, the properties of exotic nuclei have played a special role in constraining the nuclear forces. Specifically, the importance of the three-body force and  the role of the tensor force are exhibited in exotic states, such as the halo~\cite{85IT,93MVZ,13HWH} and Hoyle states~\cite{06WVO,11EEP}, the shell evolution anomaly~\cite{10TO,10TOS} and the novel magic numbers far off the $\beta$-stability~\cite{01TO,05TO,06OS}. These imply that kaonic nuclei can perhaps be regarded as candidates to probe the symmetry energy at high densities. On the other hand, kaonic nuclei themselves are fascinating systems owing to their special features, such as the breaking of pseudospin symmetry~\cite{14RYY} and the halo structure of light kaonic nuclei~\cite{Yang17}. Therefore, many theoretical works have focused on the study of kaonic nuclei~\cite{02YA,02TY,06JM,06XHZ,07DG,14RYY,Yang17}, and continuous experimental efforts have been made to search for them~\cite{05MA,06VKM,07GB,10TY,13SA,14AOT,15YI,15AF,15GA}, though the existence of kaonic nuclei has  not yet been verified. To the best of our knowledge, the relationship between kaonic nuclei and symmetry energy has seldom been investigated. Hence, we will try to construct a data-to-data correlation between the ground-state properties of kaonic nuclei and the symmetry energy in this paper under the framework of the relativistic mean-field (RMF) theory. In the RMF theory,  the interactions are mediated by mesons, which makes it very easy to incorporate a $K^-$ meson self-consistently. A characteristic of the RMF theory  is that there is a large attractive scalar and a repulsive vector, which is of great importance for interpreting the pseudospin symmetry in nuclei~\cite{Ginocchio97,Liang15}. Meanwhile, the RMF theory can not only reproduce nicely the ground-state properties for finite nuclei in the whole nuclide table~\cite{Walecka86,Ring96,Lalazissis97,Bender03}, but also  give a successful description of exotic nuclei~\cite{Lalazissis95,Meng06}. In the past, plenty of research on hyper-nuclear systems was also based on the RMF theory~\cite{94JS,85Zhang,Mares94,96Ma,06WZJ,Sun16,Sun17}. Thus, the RMF theory should be competent in our investigation on kaonic nuclei. It will be found that the strong antikaon-nucleon attraction can considerably amplify the sensitivity of the ground-state properties of kaonic nuclei to the symmetry energy at high densities.

The paper is organized as follows. In Section 2, the RMF formulas for the symmetry energy and kaonic nuclei are deduced. The numerical results are presented in Section 3, followed by a brief summary.

\section{Formalism}
To simulate different density dependence of the symmetry energy, we introduce a coupling term between the isoscalar vector and isovector vector potential, as done in Ref.~\cite{Horowitz01}. The interacting Lagrangian density for nucleons is given by
\begin{eqnarray} \label{LN}
\mathcal{L}_{int} = \bar{\psi}_B [ g_{\sigma} \sigma - g_{\omega} \gamma_{\mu} \omega^{\mu} - g_{\rho} \gamma_{\mu} \tau_3 b_0^{\mu} - e \frac{1+\tau_3}{2} \gamma_{\mu} A^{\mu}] \psi_B \nonumber\\
 - \frac{1}{3} g_2\sigma^3 - \frac{1}{4} g_3 \sigma^4 +  \frac{1}{4} c_3 (\omega_\mu \omega^\mu)^2 +4\Lambda_v g^2_\rho g^2_\omega \omega_\mu \omega^\mu b_{0\mu} b_0^\mu,
\end{eqnarray}
where $\sigma,\ \omega,\ b_0,\ A$ represent the isoscalar scalar, isoscalar vector, isovector vector, and electromagnetic field, respectively. $g_i$ ($i=\sigma,\omega,\rho$) are the corresponding coupling constants between meson fields and nucleons. $g_2$ and $g_3$ denote the strengths of the nonlinear terms of the scalar field.  $c_3$ is the parameter for the non-linear $\omega$ self-coupling term. $\Lambda_v$ is the coupling between the isoscalar vector and isovector vector potential which could be used to adjust the high-density symmetry energy.

In the mean-field approximation, we can derive the symmetry energy from the Lagrangian density Eq.~(\ref{LN}) using $E_{sym}(\rho_B)=\frac{1}{2}[\partial^2 E(\rho_B,\delta)/\partial\delta^2]|_{\delta=0}$ where $\delta$ is the isospin asymmetry. Then, we get

\begin{eqnarray}
E_{sym}(\rho_B)=\frac{1}{2}(\frac{g_\rho}{m^*_\rho})^2 \rho_B + \frac{k_F^2}{6E^*_F},
\end{eqnarray}
where the effective mass of the $\rho$ meson $m^*_\rho=\sqrt{m^2_\rho+8\Lambda _v(g_{\omega } g_{\rho}\omega _0)^2}$ and $E^*_F=\sqrt{k^2_F+M^{*2}}$ is the Fermi energy with the effective mass of the nucleon $M^*=M_B-g_\sigma \sigma$.

At a particular density $\rho'$, the symmetry energy can be approximately expanded as
\begin{eqnarray}
E_{sym}(\rho_B)=E_{sym}(\rho')+\frac{L}{3}\frac{\rho_B-\rho'}{\rho'}+\frac{\kappa_{sym}}{18}\frac{(\rho_B-\rho')^2}{\rho'^2},
\end{eqnarray}
where L and $\kappa_{sym}$ are the density slope and curvature of the symmetry energy at $\rho'$, respectively. The density slope L is defined as $L(\rho')=3\rho'\frac{\partial E_{sym}}{\partial \rho}|_{\rho=\rho'}$. Some studies have found that L and $\kappa_{sym}$ are correlated with each other~\cite{Chen11}. As $\kappa_{sym}$ is very poorly known, the determination of L with high precision could give valuable information on the symmetry energy.

To investigate kaonic nuclei, the antikaonic sector should be incorporated into the interacting Lagrangian density for nucleons Eq.~(\ref{LN}). The Lagrangian density for kaons is written as~\cite{07DG}
\begin{eqnarray}
\label{LK}
\mathcal{L}_{KN} = (\mathcal{D}_{\mu}K)^{\dag}(\mathcal{D}^{\mu}K)
 - (m^2_{K} - g_{\sigma K} m_{K} \sigma )K^{\dag}K,
\end{eqnarray}
where the covariant derivative is defined as
$\mathcal{D}_{\mu} \equiv \partial _{\mu} + i V_\mu$
with $V_\mu = g_{\omega K} \omega_{\mu} +  g_{\rho K}
b_{0\mu} + e\frac{1+\tau_3}{2}A_{\mu}$, and $K=\binom{K^+}{K^0}$, $K^\dag=(K^-,\bar{K}^0)$. In this paper, we only consider charged kaons, since kaonic nuclei are systems where the $K^-$ meson is implanted into a finite nucleus, i.e., $K=K^+$ and $K^\dag=K^-$. From this Lagrangian, it can be inferred that the $K^-N$ interaction is mediated by $\sigma,\ \omega,\ \rho$ meson fields and the electromagnetic field, which are coherently attractive, consistent with experimental data. This makes the formation of a $K^-$-nucleus bound state possible.

With the Lagrangian density for nucleons Eq.~(\ref{LN}) and $K^-$ meson Eq.~(\ref{LK}), together with the Lagrangian densities for free particles, we can obtain the equations of motion using the standard RMF treatment~\cite{Walecka86}, which can be written as


\begin{eqnarray}
&&[-i \bm{\alpha} \cdot \bm{\nabla} + \beta ( M_{B} - g_{\sigma} \sigma ) + g_{\omega} \omega + g_{\rho} \tau_3 b_0 + e\frac{1+\tau_3}{2} A]\psi_B = \epsilon \psi_B, \label{Dirac}\\
&&( -\bm{\nabla} \cdot \bm{\nabla} +  m^2_{K} -E^2_{K^-} + \Pi )K^- =0,
\label{KG_K} \\
&&( -\bm{\nabla} \cdot \bm{\nabla} + m^2_\phi )\phi = s_\phi =\left\{
\begin{aligned}{}
&g_{\sigma} \rho_s - g_2\sigma^2 - g_3\sigma^3 + g_{\sigma K}m_K K^-K^+  & \sigma, &\\
&g_{\omega} \rho_v  -c_3\omega^3 -8\Lambda_v g^2_\rho g^2_\omega \omega b^2_{0} - g_{\omega K}\rho_{K^-} & \omega, &\\
&g_{\rho} \rho_3 - 8\Lambda_v g^2_\rho g^2_\omega \omega^2 b_{0} - g_{\rho K}\rho_{K^-}, & \rho, &\\
&e \rho_p - e\rho_{K^-}, & photon, &\\
\end{aligned}
\right.
\end{eqnarray}

\noindent where $\rho_s,\ \rho_v,\ \rho_3,\ \rho_p,\ \rho_{K^-}$ are the scalar, vector, isovector, proton, and $K^-$ density, respectively. $E_{K^-}$ is the energy eigenvalue of the $K^-$ meson. $s_\phi$ is the so-called source term for mesons or photons (and $m_{photon}=0$). The $K^-$ self-energy term is expressed as
\begin{equation}
\label{SK}
\Pi = - g_{\sigma K} m_{K} \sigma  - 2E_{K^-}V - V^2,
\end{equation}
where $V = g_{\omega K} \omega +  g_{\rho K}b_{0} + \frac{1+\tau_3}{2}eA$. The parameters in Eq.~(\ref{LK}) can be fitted to the depth of the $K^-$ optical potential at normal nuclear density, where the optical potential is defined as $U_{opt}(\textbf{p},\rho_0) =
\omega(\textbf{p},\rho_0)-\sqrt{\textbf{p}^2 + m_{K}^2}$, with $\omega(\textbf{p},\rho_0)$ being the in-medium energy at saturation density. Strictly, one should consider the effects from the $K^-$ absorption by nucleons. This could be done by introducing an imaginary potential into the $K^-$ self-energy term $\Pi$ in Eq.~(\ref{SK})~\cite{06JM,06XHZ,07DG}. We have found that the stationary-state properties of  kaonic nuclei are only slightly affected by the incorporation of the imaginary part in the RMF models when the $K^-$ optical potential is around -100 MeV depth at normal nuclear density. Therefore, the imaginary part in the $K^-$ self-energy term is neglected here, as it has no impact on the conclusion in the paper.

Solving these equations in an iterative procedure, one can accurately obtain the ground-state properties of normal nuclei and kaonic nuclei. In each iteration step, we use the shooting method to obtain energy eigenvalues for nucleons and antikaons. Specifically, the energy eigenvalues are moderately adjusted to connect  the wave function smoothly at  matching points in this boundary value problem.  Note that the antikaon eigenvalue is not sensitive to the choice of the specific first-order derivative at boundaries.   Readers can refer to Ref.~\cite{14RYY} for more concrete numerical details. Then, we can construct the correlations between the structural properties of these finite nuclear systems and different density dependence of the symmetry energy.

\section{Results and discussion}
The antikaon-nucleon interaction is demonstrated to be strongly attractive  from  the data extracted from kaonic atoms~\cite{81batty,94EF,97CJB,99EF,01gal,07EF}, KN scatterings~\cite{96waas,97waas,97JSB,00AR,01AC,11AC} and heavy-ion collisions~\cite{97GQL,99WC,99FL,03AF,06WS,14ZQF}. The data from kaonic atoms and KN scattering mainly provide information on low-density $K^-N$ interactions. One needs to use some specific extrapolations to obtain the depth of the $K^-$ optical potential at normal nuclear density. However, these extrapolations cause large diversification, giving the depth at normal nuclear density ranging from -200 MeV to -50 MeV according to different models~\cite{81batty,94EF,97CJB,99EF,01gal,07EF,96waas,97waas,97JSB,00AR,01AC,11AC}. Heavy-ion collisions can directly probe the properties at appropriately produced densities and produce an almost consistent $K^-$ potential depth around -100 MeV at normal nuclear density~\cite{97GQL,99WC,99FL,03AF,06WS,14ZQF}. Thus,  we adopt -100 MeV as the $K^-$ optical potential depth at saturation density~\cite{14ZQF}, to determine the value of the unique free parameter $g_{\sigma K}$, as $g_{\rho K}$ and $g_{\omega K}$ can be determined by the SU(3) relation: $2g_{\omega K}=2g_ { \rho K}=g_ { \rho \pi}=6.04$.

\begin{table*}
\begin{center}
\caption{ The parameters $\Lambda_v$ and $g_\rho$ that simulate various density dependencies of the symmetry energy. The symmetry energy at $\rho_0$, $1.5\rho_0$ and its density slope at $0.7\rho_0$ are in units of MeV. Other unlisted parameters are the same as the TM2~\cite{94Sugahara}. \label{T1}}
\footnotesize
\begin{tabular*}{80mm}{c@{\extracolsep{\fill}}ccccc}
\toprule  $\Lambda_v$ & $g_\rho$ & L($0.7\rho_0$) & $E_{sym}$($\rho_0$) & $E_{sym}$($1.5\rho_0$) \\
\hline
0.0   & 9.357 & 106.53 & 35.90 & 55.19\\
0.005 & 9.778  & 97.65 & 34.55 & 49.81\\
0.010 & 10.264 & 88.79 & 33.39 & 46.17\\
0.015 & 10.830 & 79.92 & 32.37 & 43.53\\
0.020 & 11.502 & 71.05 & 31.47 & 41.54\\
0.025 & 12.316 & 62.19 & 30.68 & 39.97\\
0.030 & 13.332 & 53.31 & 29.96 & 38.71\\
\hline
\end{tabular*}
\vspace{0mm}
\end{center}
\end{table*}

\begin{figure}
\begin{center}
\includegraphics[width=6.5cm]{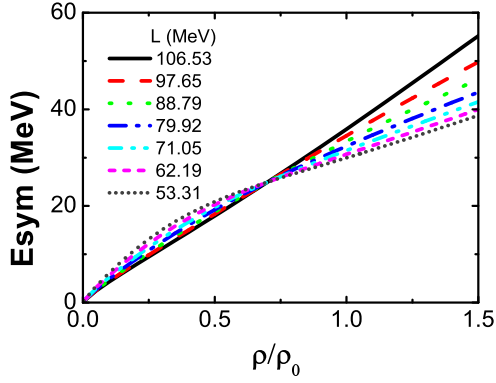}
\caption{\label{Fesym} (Color online) The symmetry energy as a function of nuclear matter density with various slopes (L) at 0.7$\rho_0$ obtained from adjusting parameter $\Lambda_v$ and $g_\rho$ based on TM2, see Table~\ref{T1}.}
\end{center}
\end{figure}

We adopt the RMF parameter set TM2 as a starting point in this work, since TM2 was initially designed for light nuclei~\cite{94Sugahara}. The different density dependence of the symmetry energy can be obtained by adjusting the parameters $\Lambda_v$ and $g_\rho$ in  Eq.~(\ref{LN}) with a fixed a symmetry energy of 24.93MeV at $\rho=0.7\rho_0$ (the same as TM2 and $\rho_0=0.132fm^{-3}$), similar to what was done in Ref.~\cite{Horowitz01}. This simple treatment produces a nearly constant binding energy per nucleon for oxygen isotopes when $\Lambda_v$ is changed. However, the density dependence of the symmetry energy diversifies considerably at high densities due to different coupling strengths between isoscalar vector and isovector vector field, as shown in Fig.~\ref{Fesym}. When $\Lambda_v$ changes from 0 to 0.03, the symmetry energy becomes softer at high densities and the density slope of the symmetry energy at 0.7$\rho_0$ varies from 106.53 MeV to 53.31 MeV.  The specific parameters and properties are listed in Table~\ref{T1}.

\begin{figure}
\begin{center}
\includegraphics[width=6.5cm]{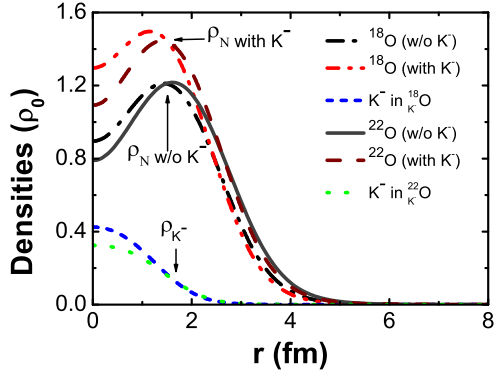}
\caption{\label{Fdens} (Color online) The nuclear density and $K^-$ density distribution in normal nuclei ($^{18}O$, $^{22}O$) and kaonic nuclei ($^{18}_{K^-}O$, $^{22}_{K^-}O$) with the parameter set TM2. The corresponding density distributions with various $\Lambda_v$ are very similar to the one with the TM2, thus they are not shown here for simplicity. The  ``w/o $K^-$"   denotes  normal nuclei without $K^-$ and the ``with $K^-$" denotes the corresponding kaonic nuclei.   }
\end{center}
\end{figure}

Using the parameters in Table~\ref{T1}, we calculate the ground-state properties of oxygen isotopes, including the density distributions, nuclear potentials and the single-particle energy.  In Fig.~\ref{Fdens}, we display the nuclear density and $K^-$ density distributions in $^{18}O$ and $^{22}O$ systems. As one can see, in the core of kaonic nuclei, the nuclear density is largely increased by the implanted $K^-$ meson. This is the so-called shrinkage effect, which may be constructive for producing cold high-density matter in the laboratory. In fact, it is the strongly attractive interaction between nucleons and $K^-$ meson that makes the compression in the core of a nucleus possible, resulting in this shrinkage effect. Indeed, the increase of the nuclear density in kaonic nuclei almost coincides with the $K^-$ density distribution ,which is mainly stacked in the core region of kaonic nuclei.

\begin{figure}
\begin{center}
\includegraphics[width=6.0cm]{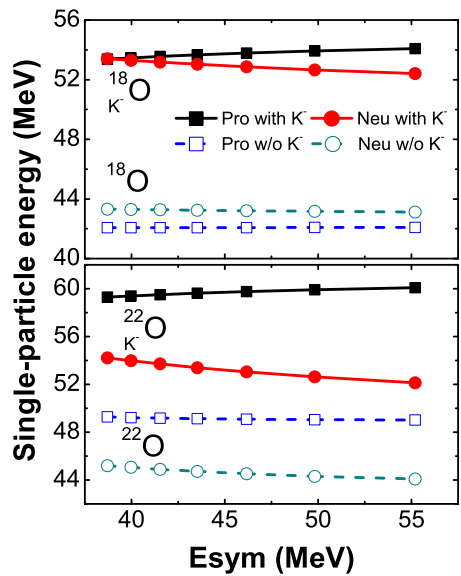}
\caption{\label{Fsp} (Color online) The single-particle energies of the $1S_{1/2}$ state in $^{18}O$ and $^{22}O$ systems as a function of the symmetry energy at 1.5$\rho_0$. The ``Pro" and ``Neu" labels denote the $1S_{1/2}$ state for protons and neutrons, respectively.}
\end{center}
\end{figure}

\begin{figure}
\begin{center}
\includegraphics[width=7cm]{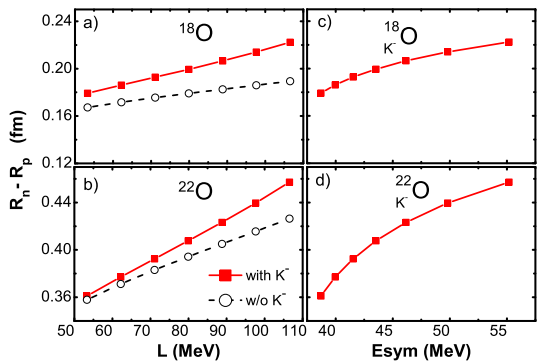}
\caption{\label{Flesymskin} (Color online) The neutron skin thickness in $^{18}O$ and $^{22}O$ systems as a function of the slope of symmetry energy at 0.7$\rho_0$ (left-hand panels) and as a function of the symmetry energy at 1.5$\rho_0$ (right-hand panels).}
\end{center}
\end{figure}

\begin{figure}
\begin{center}
\includegraphics[width=7.0cm]{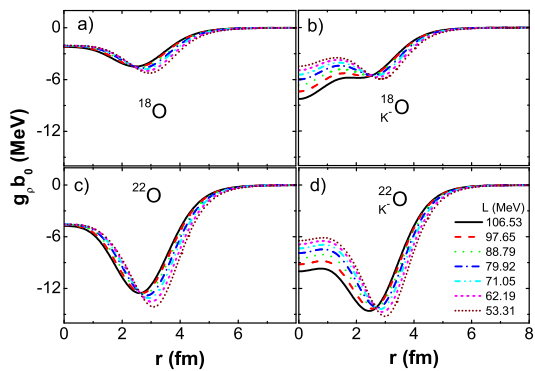}
\caption{\label{Fpot} (Color online) The isovector potential in $^{18}O$ and $^{18}_{K^-}O$, $^{22}O$ and $^{22}_{K^-}O$ as a function of radius. The $L$ value marked is obtained at 0.7$\rho_0$.}
\end{center}
\end{figure}

Accompanying the shrinkage effect, the inner nuclear states in kaonic nuclei are considerably affected by the embedding of the $K^-$ meson. Shown in Fig.~\ref{Fsp} is the single-particle energy of the $1S_{1/2}$ state for protons and neutrons as a function of symmetry energy at 1.5$\rho_0$ which can be adjusted through different $\Lambda_v$ in Table~\ref{T1}. The kaonic nuclei have an average core density around 1.5$\rho_0$, and the correlation between the $1S_{1/2}$ state energy in the core region and the symmetry energy at  1.5$\rho_0$ can directly probe the variation of the symmetry energy at supranormal nuclear density.  We can see that the single-particle energies of $1S_{1/2}$ states for kaonic nuclei (solid curves) are clearly larger than those for the corresponding normal nuclei (dashed curves), i.e., the nucleons in the $1S_{1/2}$ state of kaonic nuclei are more tightly bound. This is a straightforward consequence of the strong antikaon-nucleon attraction that would change the potentials for nucleons in nuclei. Noticeably, the difference in the single-particle energy between protons and neutrons in kaonic nuclei becomes increasingly large with the increase of high-density symmetry energy, while it is much less obvious in normal nuclei. This actually establishes a direct signal to sensitively probe the high-density symmetry energy.
The rms quantities can be calculated using the density distributions as weight functions. Here, we focus on the neutron and proton rms radii and define the neutron skin thickness as the difference between the neutron and proton rms radii. The results for the neutron skin thickness in neutron-rich $^{18}O$ and $^{22}O$ systems are plotted in Fig.~\ref{Flesymskin}. As one can see from the left-hand panels in Fig.~\ref{Flesymskin}, the relationship between the neutron skin thickness and the density slope of the symmetry energy at $0.7\rho_0$ is nearly linear in both $^{18}O$ and $^{22}O$ systems. The neutron skin thickness in kaonic nuclei (solid lines) is larger than that in normal nuclei (dashed lines) due to the shrinkage effect that occurs in the inner states of kaonic nuclei. The gradient of the neutron skin thickness with increasing density slope of the symmetry energy in $^{18}_{K^-}O$ is moderately larger than that in normal nucleus $^{18}O$. Strikingly, the neutron skin thickness in $^{22}_{K^-}O$ is obviously more sensitive to the density slope of the symmetry energy, as compared to that of $^{22}O$. To further understand the role of the $K^-$ meson in affecting the sensitivity of the neutron skin thickness  to the difference in the symmetry energy, we display in the right-hand panels of Fig.~\ref{Flesymskin} the neutron skin thickness of $^{18}_{K^-}O$ and $^{22}_{K^-}O$ as a function of the symmetry energy at 1.5$\rho_0$. Since this density appears to be close to the average density of the core of kaonic nuclei,  the consequence of the symmetry energy   in the core region could also be reflected to the neutron skin.  It can be clearly seen from panel (d) in Fig.~\ref{Flesymskin} that the neutron skin thickness in $^{22}_{K^-}O$ changes rather rapidly  with increasing symmetry energy at 1.5$\rho_0$. The smaller variation observed for  $^{18}_{K^-}O$ in panel (c), compared to that of $^{22}_{K^-}O$, is attributed to the smaller isospin asymmetry. This can be also seen by comparing the curves for normal nuclei $^{22}O$ and  $^{18}O$, as shown in  Fig.~\ref{Flesymskin}. Nevertheless, the sensitivities of the neutron skin thickness to the isospin effect and symmetry energy are both magnified by the embedding of the $K^-$ meson.
To reveal the physics behind these phenomena, we plot in Fig.~\ref{Fpot} the isovector  potential in $^{18}O$ and $^{18}_{K^-}O$, $^{22}O$ and $^{22}_{K^-}O$ as a function of radius. In normal nucleus $^{18}O$, the isovector potential is very small, in accordance with its small isospin asymmetry. The variation of the isovector potential with different density dependence of the symmetry energy is insignificant in $^{18}O$, responsible for a relatively slow change of the neutron skin thickness (see Fig.~\ref{Flesymskin}). However, the isovector potential in the core of $^{18}_{K^-}O$ undergoes an obvious change for various symmetry energies, in sharp contrast to the case in $^{18}O$. Thus, the embedment of $K^-$ meson makes the effect of different symmetry energy on the isovector potential more distinguishable. The situation for $^{22}O$ systems in the lower panels of Fig.~\ref{Fpot} is very similar to that of $^{18}O$ systems. However, the isovector potential in $^{22}O$ is much larger than that in $^{18}O$, since $^{22}O$ possesses a larger isospin asymmetry.
Actually, the symmetry energy effect on the isovector potentials in $^{22}_{K^-}O$ is amplified by both the implanted $K^-$ meson and the larger isospin asymmetry.
Specifically, the potential in the central region of $^{22}_{K^-}O$ has a 55\% variation as the density slope L changes from  53.31 MeV to 106.53 MeV. Hence, the implantation of the $K^-$ meson greatly magnifies the difference in the isovector  potential that is induced by the various density dependencies of the symmetry energy, responsible for the larger changes in the neutron skin thickness of kaonic nuclei.
We have also conducted a series of investigations based on the RMF parameter set NL-SH~\cite{NLSH} and FSUGold~\cite{FSUGold}. It is found that the conclusions are qualitatively the same. These results indicate that light kaonic nuclei can be used as  candidates to constrain the high-density symmetry energy either with appropriate theoretical approaches or by performing experiments. The photo-nucleus or pion-nucleus reactions~\cite{Kaiser97,14AOT}, e.g., $\gamma$ + $^{22}$O $\rightarrow$ $K^+$  +  $^{22}_{K^-}$O or $\pi^-$ + $^{22}$F $\rightarrow$ $K^{*+}$ + $^{22}_{K^-}$O, may be used to produce kaonic nuclei and then directly measure the radius information via the strong interaction between the nucleons and the outgoing kaons in these reactions.

\section{Summary}
\label{summary}
In this paper, we have investigated the correlation between the ground-state properties, especially the neutron skin thickness, of kaonic oxygen isotopes and the symmetry energy in the framework of the RMF theory. By introducing a coupling between the isoscalar vector and isovector vector fields, we can simulate different density dependencies of the symmetry energy. We have directly established  a sensitive correlation between  the ground-state properties of kaonic oxygen isotopes and the symmetry energy at supranormal density.  It is found that
the sensitivity of the neutron skin thickness to the low-density slope of the symmetry energy is greatly amplified in the corresponding kaonic nuclei. These sensitive relationships are established upon the fact that   the isovector potential in the central region, where a high density core is formed due to the embedding of the $K^-$ meson, becomes very sensitive to the variation of the symmetry energy. These amplified sensitivities could be instructive for constraining the high-density features of the symmetry energy.

\section*{Acknowledgements}
The work was supported in part by the National Natural Science
Foundation of China under Grant Nos. 11775049 and 11275048 and the
China Jiangsu Provincial Natural Science Foundation under Grant
No. BK20131286.

\clearpage

\end{document}